\documentclass[%
reprint,
 amsmath,amssymb,
 aps,
]{revtex4-2}

\usepackage{dcolumn}
\usepackage{bm}
\usepackage{hyperref, soul}
\hypersetup{linktocpage,,colorlinks=true}
\usepackage{hyperref, soul}
\usepackage{physics}
\usepackage{subcaption}
\usepackage{graphicx}
\usepackage{cleveref}
\usepackage{tikz}
\usepackage{amsmath, amssymb}
\usepackage{comment}
\usepackage{mathtools}

\newcommand{\panel}[3]{%
  \begin{tikzpicture}[inner sep=0pt, outer sep=0pt]
    \node[anchor=south west, inner sep=1.5pt] (img) at (0,0)
      {\includegraphics[width=\dimexpr #2-3pt\relax]{#1}}; 
    \node[anchor=north west, font=\bfseries, fill=white, rounded corners=0pt, inner sep=0pt]
      at ([xshift=0pt,yshift=-0.1pt]img.north west) {#3};
  \end{tikzpicture}%
}

\newcommand{\panelH}[3]{
	\begin{tikzpicture}[inner sep=0pt, outer sep=0pt,
	baseline=(current bounding box.north)]
	\node[anchor=south west, inner sep=1.5pt] (img) at (0,0)
	{\includegraphics[height=\dimexpr #2-3pt\relax,keepaspectratio]{#1}};
	\path ([yshift=0.6ex]img.north west) -- ([yshift=0.6ex]img.north east);
	\node[anchor=north west, font=\bfseries, fill=white,
	inner sep=0.6pt, text height=1.3ex, text depth=0.25ex]
	at ([yshift=-0.4ex]img.north west) {#3};
	\end{tikzpicture}%
}

\makeatletter
\newcommand{\SM@oldaddcontentsline}{} 

\newcommand{\SMtocredirectON}{%
  \let\SM@oldaddcontentsline\addcontentsline
  \renewcommand{\addcontentsline}[3]{%
    \def\SM@file{##1}%
    \def\SM@toc{toc}%
    \ifx\SM@file\SM@toc
      \SM@oldaddcontentsline{smtoc}{##2}{##3}%
    \else
      \SM@oldaddcontentsline{##1}{##2}{##3}%
    \fi
  }%
}

\newcommand{\SMtocredirectOFF}{%
  \let\addcontentsline\SM@oldaddcontentsline
}

\newcommand{\SMtableofcontents}{%
  \begingroup
    \setcounter{tocdepth}{2}
    \par\noindent{\bfseries Contents}\par\medskip
    \@starttoc{smtoc}%
  \endgroup
}
\makeatother

\begin{document}

\preprint{APS/123-QED}

\title{Beyond Calabrese–Cardy Scaling:\\ Exceptional-Point Sensitivity and the de Sitter RT Surface}

\author{Kuang-Hung Chou}
 \email{nagisa.7256.960247@gmail.com}
\affiliation{%
 Department of Physics, National Tsing Hua University, Hsinchu 30013, Taiwan
}%

\date{\today}

\begin{abstract}
    Entanglement entropy at one-dimensional criticality typically follows
    the Calabrese--Cardy scaling.
    In the non-Hermitian critical free-fermion systems studied here,
    however, we find an additional interval-independent contribution
    $S_{\rm res}=\log(\Delta L)$, which allows the entropy to remain
    sensitive to energy gaps below the finite-size scale, $\Delta<1/L$.
    We refer to this term as the residual entropy.
    To understand its infrared origin, we examine finite-depth non-unitary continuous multiscale entanglement renormalization on a ring. Terminating the critical RG circuit before its singular final layer leaves a two-site infrared state. We calculate the entropy of this state directly from its correlation matrix and recover the same \(\log(\Delta L)\) dependence. The emergent de Sitter geometry provides a complementary geometric interpretation: extremal curves associated with different boundary intervals extend to a common infrared endpoint, suggesting how the entanglement retained there can contribute equally to their entropies.
\end{abstract}

\maketitle


\emph{Introduction.}---Entanglement entropy is a standard probe of one-dimensional quantum criticality and a central quantity in holography.  
For the ground state of a \(1+1\)-dimensional conformal field theory (CFT), the entanglement entropy of a single interval is fixed by conformal symmetry and exhibits the characteristic Calabrese--Cardy logarithm~\cite{Holzhey1994GeometricAR,PasqualeCalabrese_2004,Calabrese_2009}. 
In holographic theories, the same scaling follows from the Ryu--Takayanagi (RT) formula~\cite{PhysRevLett.96.181602,VeronikaE.Hubeny_2007},
where the entropy is computed from a bulk extremal curve anchored at the endpoints of the boundary interval. 
A related geometric picture arises in entanglement renormalization approaches, including the multiscale entanglement renormalization ansatz (MERA) and its continuum version, continuous MERA (cMERA), which organize quantum entanglement along the renormalization direction~\cite{PhysRevLett.101.110501,PhysRevD.86.065007,Swingle2012ConstructingHS,PhysRevLett.110.100402,Nozaki2012,PhysRevB.94.075124}.

Entanglement diagnostics have also been extended to non-unitary systems realized in non-Hermitian (nH) lattice models.
Within the biorthogonal framework~\cite{Brody_2014}, non-Hermitian critical chains can still exhibit the Calabrese--Cardy form with a non-unitary central charge~\cite{PhysRevResearch.2.033069,SciPostPhys.12.6.194,chou2026ptsymmetryenrichednonunitarycriticality,n578-ljd5,PhysRevB.107.205153}.
On the holography side, recent developments in de Sitter (dS) holography, pseudoentropy, and timelike entanglement have revealed qualitative differences between extremal surfaces in dS spacetime and the RT surfaces in anti-de Sitter (AdS) spacetime~\cite{Hikida2022,PhysRevLett.130.031601,Doi2023, PhysRevD.107.126004,PhysRevD.109.086009,PhysRevD.103.026005}.
An analogous distinction arises in entanglement renormalization, where a non-unitary extension of cMERA turns the RG direction timelike and gives rise to a de Sitter spacetime~\cite{chou2026emergentsitterspacenonunitary}.

Non-Hermitian critical systems with exceptional points (EPs) exhibit a non-diagonalizable structure associated with logarithmic conformal field theory (LCFT)~\cite{GURARIE1993535,KAUSCH2000513,io2026nonhermitianfreefermioncriticalsystems}, and their entanglement entropy can become sensitive to how the EP singularity is regulated~\cite{PhysRevResearch.2.033069,SciPostPhys.12.6.194}.
Previous analyses have primarily focused on the leading Calabrese–Cardy scaling.
Here we retain this sensitivity and consider the case in which the singularity is lifted by a small physical energy gap \(\Delta\), showing that it produces an additional contribution beyond the leading critical term.

\begin{figure}[t]
	\centering
	\begin{minipage}[b]{0.53\linewidth}
		\panelH{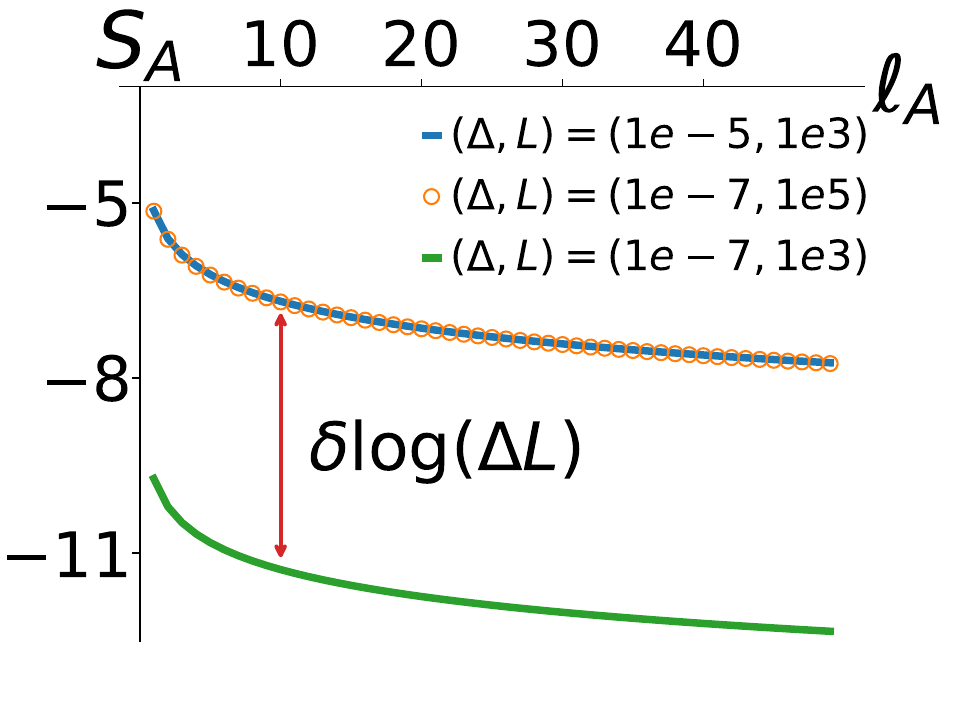}{3.4cm}{(a)}
	\end{minipage}
	\begin{minipage}[b]{0.45\linewidth}
		\panelH{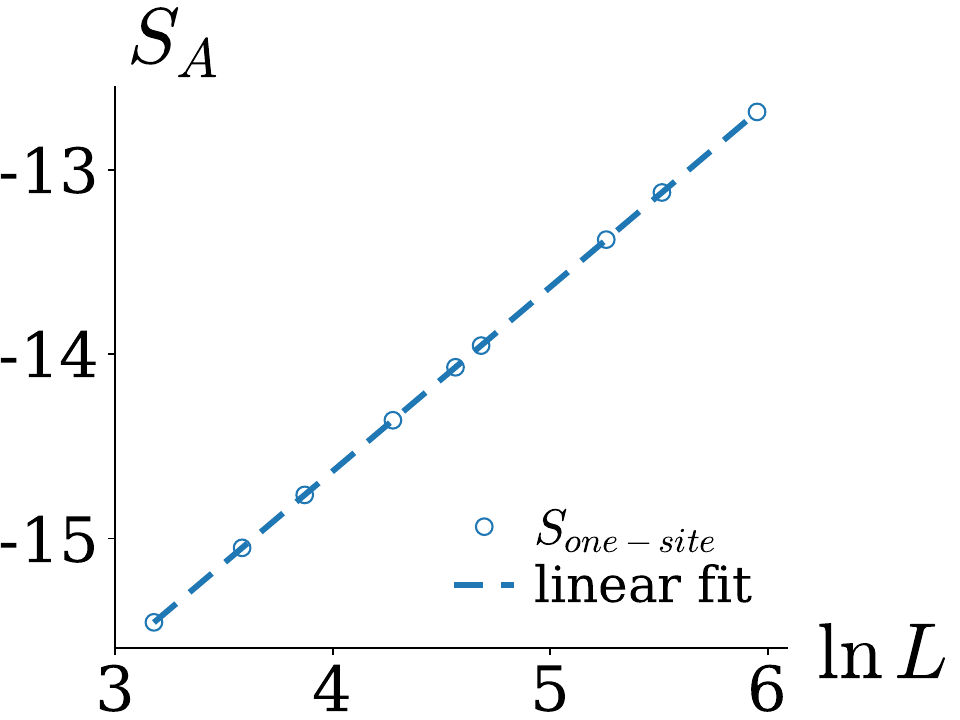}{2.9cm}{(b)}
	\end{minipage}
	\begin{minipage}[b]{0.45\linewidth}
		\panelH{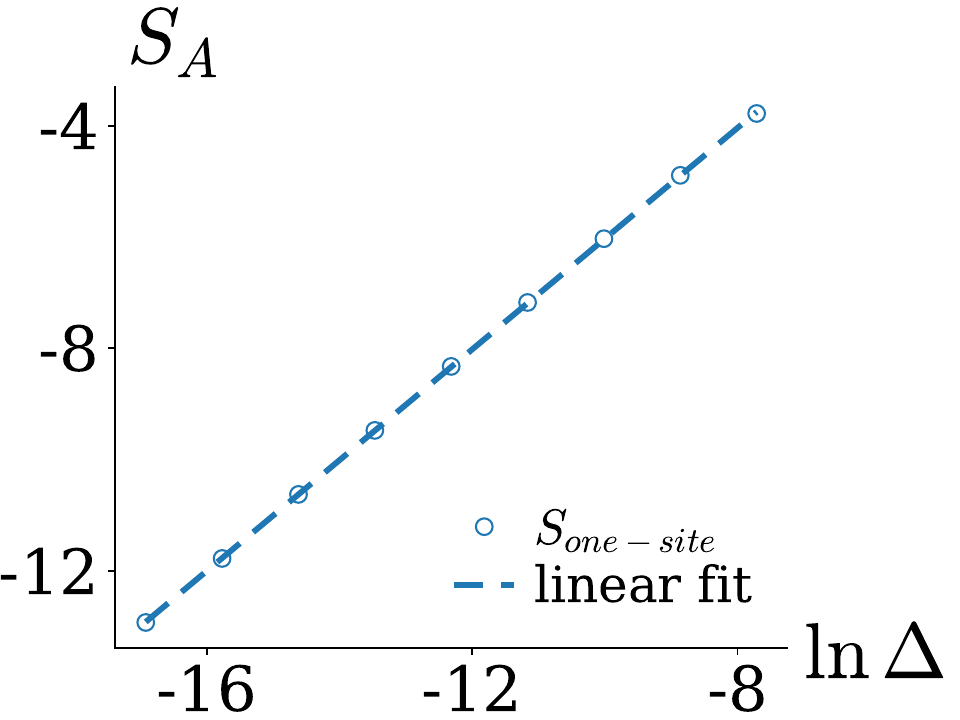}{2.9cm}{(c)}
	\end{minipage}
    \begin{minipage}[b]{0.45\linewidth}
		\panelH{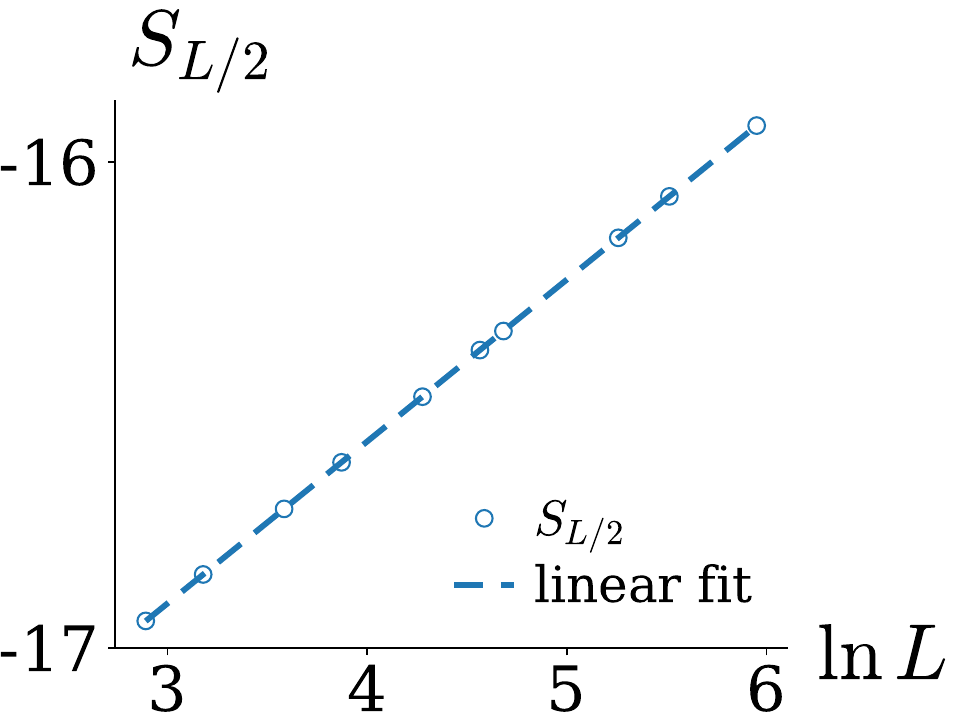}{2.9cm}{(d)}
	\end{minipage}
	\caption{
    Microscopic scaling of the residual entropy in the two-legged
    nH SSH model.
    (a) Entanglement entropy as a function of subsystem size \(\ell_A\).
    Curves with the same \(\Delta L\) collapse, while changing
    \(\Delta L\) produces an interval-independent vertical shift
    \(\delta\log(\Delta L)\).
    (b) One-site entropy as a function of \(\log L\) at fixed
    \(\Delta=10^{-8}\), with a fitted slope \(0.999\).
    (c) One-site entropy as a function of \(\log\Delta\) at fixed
    \(L=100\), with a fitted slope \(0.995\).
    (d) Half-chain entropy as a function of \(\log L\) at fixed
    \(\Delta=10^{-8}\), with a fitted slope \(0.333\), consistent with
    \(S_{L/2}=\frac{1}{3}\log L+\mathrm{const.}\) for \(c=-2\).
    Dashed lines denote linear fits.
    }
	\label{fig:residual}
\end{figure}

Concretely, for nH free fermions near an EP with a small physical energy gap \(\Delta\), the entropy of an interval of length \(\ell\) on a ring of length \(L\), in the regime \(\Delta L\ll1\),  takes the form
\begin{equation} 
S_A(\ell)= \underbrace{\frac{c}{3}\log\!\Big[\frac{L}{\pi}\sin\!\Big(\frac{\pi \ell}{L}\Big)\Big]}_{\textbf{Critical Scaling}} +\!\!\!\! \underbrace{\log(\Delta L)\vphantom{\Big[\frac{L}{\pi}\sin\!\Big(\frac{\pi \ell}{L}\Big)\Big]}}_{\textbf{Residual Entropy}} \!\!\!\! +\,\,\,s_0 , 
\label{eq:intro_decomp} 
\end{equation} 
where \(c=-2\) is the non-unitary central charge and \(s_0\) is a nonuniversal constant.
We call the additional interval-independent contribution \(S_{\rm res}\equiv\log(\Delta L)\) the \emph{residual entropy}.
Data at fixed \(\Delta L\) collapse onto the same curve, whereas changing \(\Delta L\) produces a vertical shift of the Calabrese--Cardy scaling form, as shown in Fig.~\ref{fig:residual}(a).

The separate dependences on the gap $\Delta$ and the system size $L$,
shown in Fig.~\ref{fig:residual}(b)--(d), confirm Eq.~\eqref{eq:intro_decomp}.
The numerical parameters and the procedure used to open a small gap near the EP are given in Sec. I of the SM~\cite{SM}.
The residual term remains visible even for a one-site subsystem, which
retains the dependence on both $L$ and $\Delta$, as shown in
Fig.~\ref{fig:residual}(b) and (c).
The persistence of this global \(L\) dependence in a one-site subsystem points to an infrared origin of the residual contribution.
The resulting sensitivity to gaps below the finite-size scale,
$\Delta<1/L$, is a distinctive feature of the non-Hermitian critical
systems studied here and is absent in their unitary counterparts~\cite{PasqualeCalabrese_2004}.

To investigate the infrared origin of the residual entropy, we turn to finite-depth non-unitary entanglement renormalization. We first review the emergent dS geometry and then show that terminating the RG circuit before its singular final layer leaves a two-site IR state, whose entropy reproduces the \(\log(\Delta L)\) dependence. The dS extremal curves provide a complementary geometric picture: they extend to a common IR endpoint for different boundary intervals, suggesting both sensitivity to infrared information and an interval-independent contribution.
This identifies the additional term as the long-range entanglement left
in the IR state after finite-depth disentangling, giving the residual
entropy its physical meaning.

\emph{De Sitter spacetime from non-unitary cMERA.}---To explore the physical origin of the residual entropy, we first recall the cMERA construction of Ref.~\cite{chou2026emergentsitterspacenonunitary}, where the entanglement structure of nH free fermions gives rise to a dS spacetime.  
We use the two-legged non-Hermitian Su--Schrieffer--Heeger (SSH) model
as an example; results for other non-Hermitian free-fermion models are
given in Supplemental Material (SM)~\cite{SM}.  Its Hamiltonian takes the
form \(H=\sum_k\Psi_k^\dagger h(k)\Psi_k\), with
\(\Psi_k=(c_{k,1},c_{k,2})^T\).
The single-particle Hamiltonian is
\begin{equation}
h(k)=
\begin{pmatrix}
iu & v_k \\
v_k & -iu
\end{pmatrix},
\qquad
v_k = v-w\cos k ,
\end{equation}
with energy gap \(\Delta=\sqrt{(|v|-|w|)^2-u^2}\). 
The right and left single-particle eigenstates can then be written in terms of the complex angle
\(\phi(k)=\arctan [v_k/(iu)]\). 

The cMERA circuit progressively disentangles and coarse-grains the state along the RG flow from the ultraviolet (UV) boundary at \(u=0\) toward the IR reference state \(u\to-\infty\).  
Along the RG coordinate \(u\leq0\), coarse-graining rescales momenta as \(k\to ke^{-u}\), while the disentangler of strength \(g(u)\) acts on modes whose rescaled momentum remains below the cutoff \(\pi\).  
A mode \(k\) is therefore disentangled for \(u\geq\ln(|k|/\pi)\) and is coarse-grained out when its rescaled momentum reaches the cutoff.  
The accumulated rotation from its physical band angle \(\phi(k)/2\) to the common IR value \(\phi(\pi)/2\) then satisfies
\begin{equation}
\frac{\phi(k)}{2}-\frac{\phi(\pi)}{2}
=
\int_{\ln(|k|/\pi)}^0 g(s)\,ds .
\label{eq:cmera_rotation}
\end{equation}
At the gapless critical point, the band angle becomes logarithmically singular for long-wavelength modes near the \(k=0\) exceptional point, \(\phi(k)\sim i\log k\).
Differentiating Eq.~\eqref{eq:cmera_rotation} and using \(k=\pi e^u\) gives
\begin{equation}
g(u)
=
-\frac{k}{2}\partial_k\phi(k)\bigg|_{k=\pi e^u}
\quad\Longrightarrow\quad
g(-\infty)=-\frac{i}{2},
\end{equation}
so the IR disentangler approaches an imaginary constant.  
The radial component computed from the information metric is proportional to \(g^2(u)\).  
Since \(g^2(u)<0\), the RG direction becomes timelike.
Including the spatial rescaling along the RG direction gives
\begin{equation}
ds^2
=
-\frac{1}{4}du^2
+
\pi^2e^{2u}dx^2 ,
\end{equation}
which is de Sitter spacetime in flat-slicing coordinates.  
The full non-unitary cMERA construction and the derivation of the information metric are given in Ref.~\cite{chou2026emergentsitterspacenonunitary}.

\begin{figure}[t]
    \centering
    \begin{minipage}[b]{0.48\linewidth}
        \panel{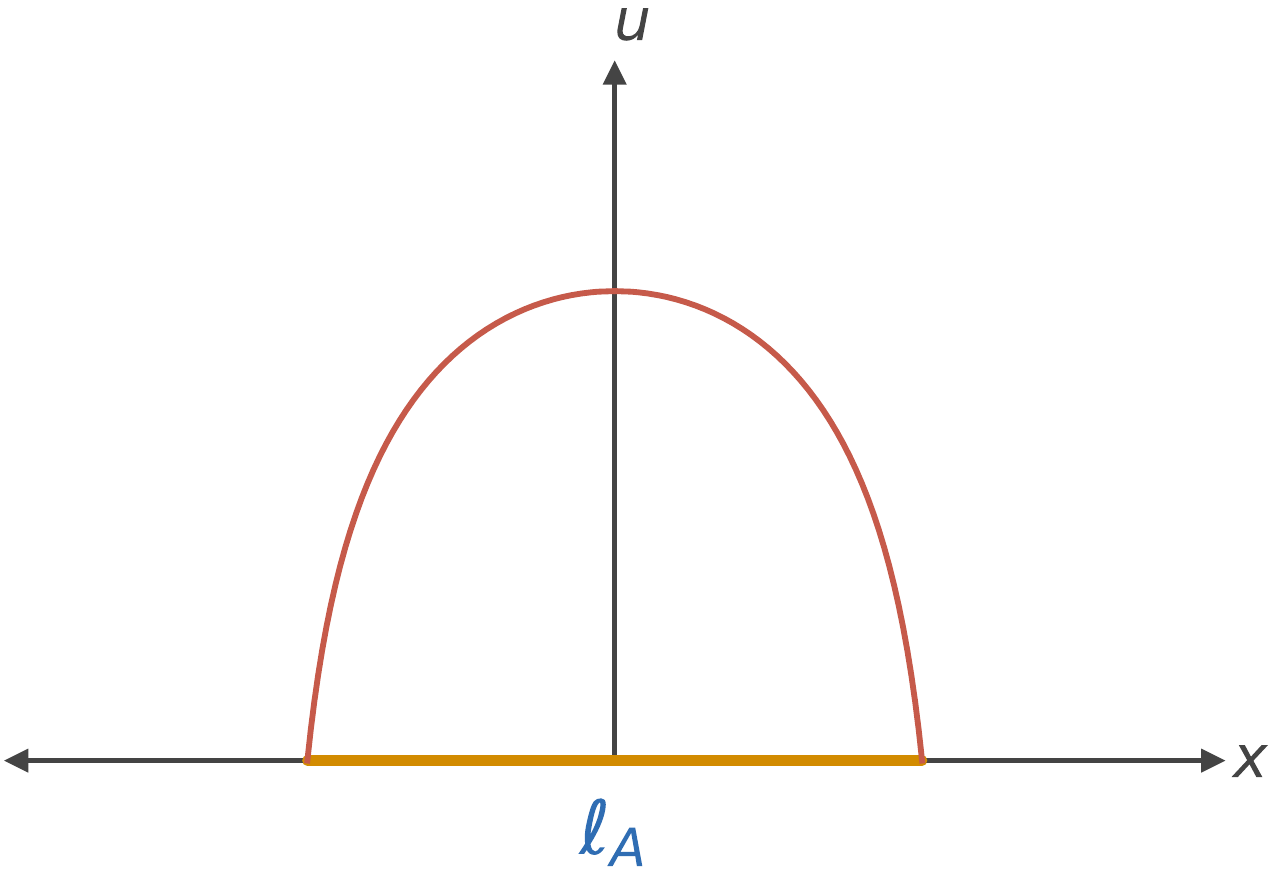}{\linewidth}{(a)}
    \end{minipage}
    \begin{minipage}[b]{0.48\linewidth}
        \panel{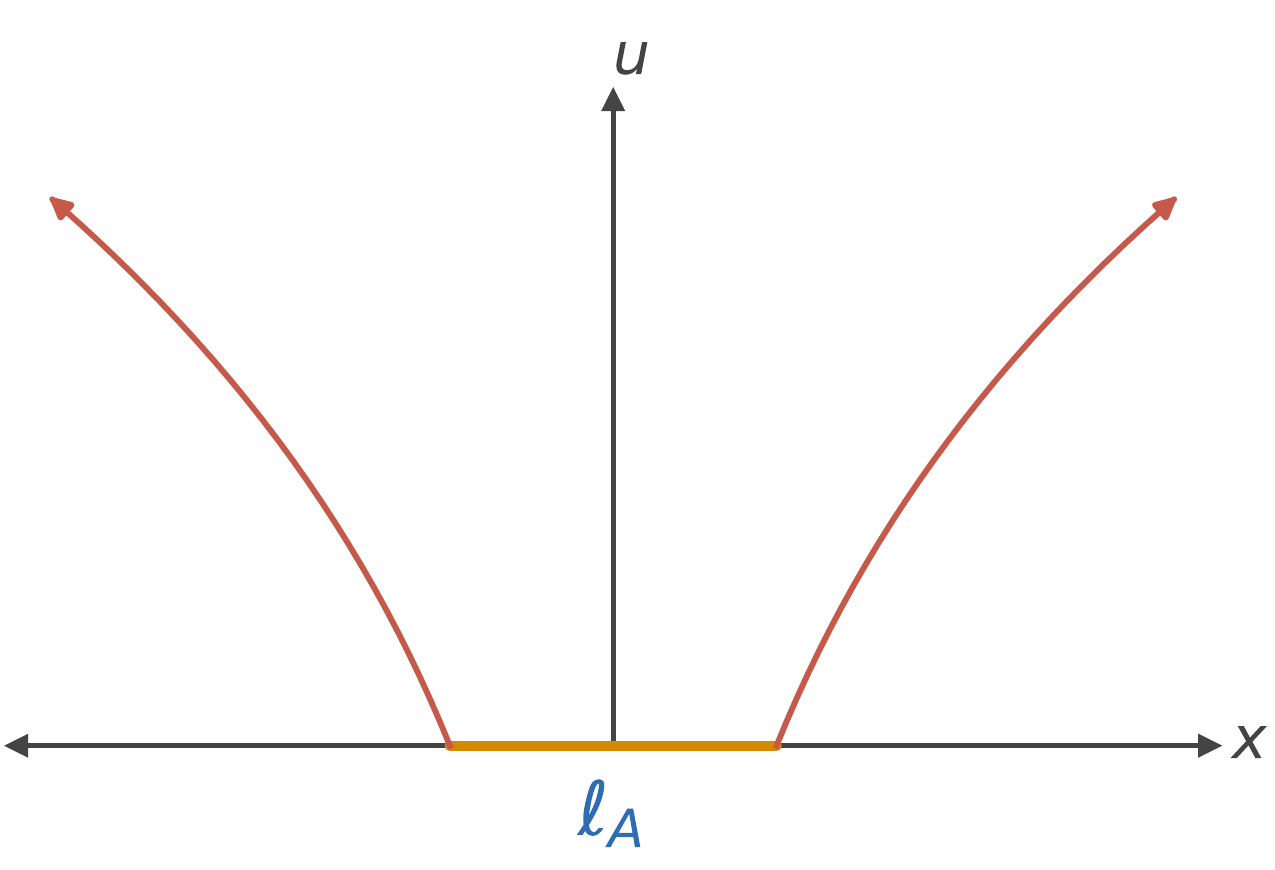}{\linewidth}{(b)}
    \end{minipage}
    \caption{
    Schematic extremal surfaces in the emergent geometries.
    (a) In AdS, the boundary-to-boundary RT surface connects the two
    endpoints of the interval \(A\).
    (b) In dS spacetime, the timelike extremal geodesics extend
    toward past infinity, corresponding to the IR endpoint of the
    RG circuit.
    }
    \label{fig:geo}
\end{figure}

Because the RG direction is timelike, the corresponding extremal curves differ qualitatively from those in AdS. In AdS, the entanglement entropy of an interval is computed by a boundary-to-boundary RT geodesic. In the emergent dS geometry, the timelike extremal curves instead extend toward past infinity, corresponding to the IR endpoint of the RG circuit. This distinction suggests that infrared degrees of freedom may contribute to the entanglement entropy of every boundary interval; see Fig.~\ref{fig:geo}.

\emph{Finite-depth RG and the remaining IR state.}---The infinite-depth cMERA circuit reviewed above contains no explicit system-size scale and therefore cannot by itself produce the dependence on \(\Delta L\).  
We now consider the RG flow on a ring of \(L\) sites, for which the circuit terminates after a finite number of layers.  
For \(L=2^N\), successive coarse-graining by a factor of two leaves a single site after \(N=\log_2L\) layers.  
It is convenient to shift the RG coordinate so that \(u=\log L\) at the UV boundary and \(u=0\) at the IR endpoint.  
A representative rescaling that implements this termination is
\begin{equation}
k\mapsto k\,\frac{\sinh\log L}{\sinh u}.
\label{eq:sinh_rescaling}
\end{equation}
Away from the endpoint, this profile reproduces the exponential momentum rescaling of the infinite-depth circuit, while its behavior at \(u=0\) implements the final collapse to a single site.

For a general finite-depth rescaling \(k\mapsto k/f(u)\), the same accumulated band rotation fixes the modified disentangler as
\begin{equation}
\frac{\phi(k)}{2}-\frac{\phi(\pi)}{2}
=
\int_{f^{-1}(k/\pi)}^{\log L}\!\tilde g(s)\,ds .
\end{equation}
Differentiating with respect to \(k\) and then substituting \(k=\pi f(u)\) yields the finite-depth disentangler
\begin{equation}
\tilde g(u)
=
-\frac{\pi f'(u)}{2}\,
\partial_k\phi(k)\Big|_{k=\pi f(u)}
=
\frac{f'(u)}{f(u)}\,g(\tilde u),
\label{eq:finite_entangler}
\end{equation}
where \(\tilde u\equiv \ln f(u)\) relates the finite-size circuit back to the infinite-size cMERA parameter.

For the rescaling in Eq.~\eqref{eq:sinh_rescaling}, \(f(0)=0\).
Since the critical disentangler remains nonzero in the IR, Eq.~\eqref{eq:finite_entangler} gives
\begin{equation}
f(0)=0
\text{ with }
g(-\infty)\neq0
\quad\Longrightarrow\quad
\widetilde g(0)\rightarrow\infty .
\label{eq:singular_endpoint}
\end{equation}
The corresponding radial metric, \(\widetilde g_{uu}=\widetilde g^{\,2}(u)\), therefore diverges at the IR endpoint.  
Thus, completing the finite-depth flow to a fully disentangled one-site state produces a singular final RG layer.  
This is the obstruction associated with long-range entanglement that cannot be removed by a finite-depth local circuit.

To avoid this singular final layer, we terminate the finite-depth flow one layer earlier. Since each RG layer halves the number of sites, this leaves a two-site IR state on the ring rather than a one-site product state. The two-site IR state is generally entangled and retains the long-wavelength degrees of freedom that cannot be removed without passing through the singular final RG layer.

\begin{figure}
    \centering
    \begin{minipage}[b]{0.48\linewidth}
        \panel{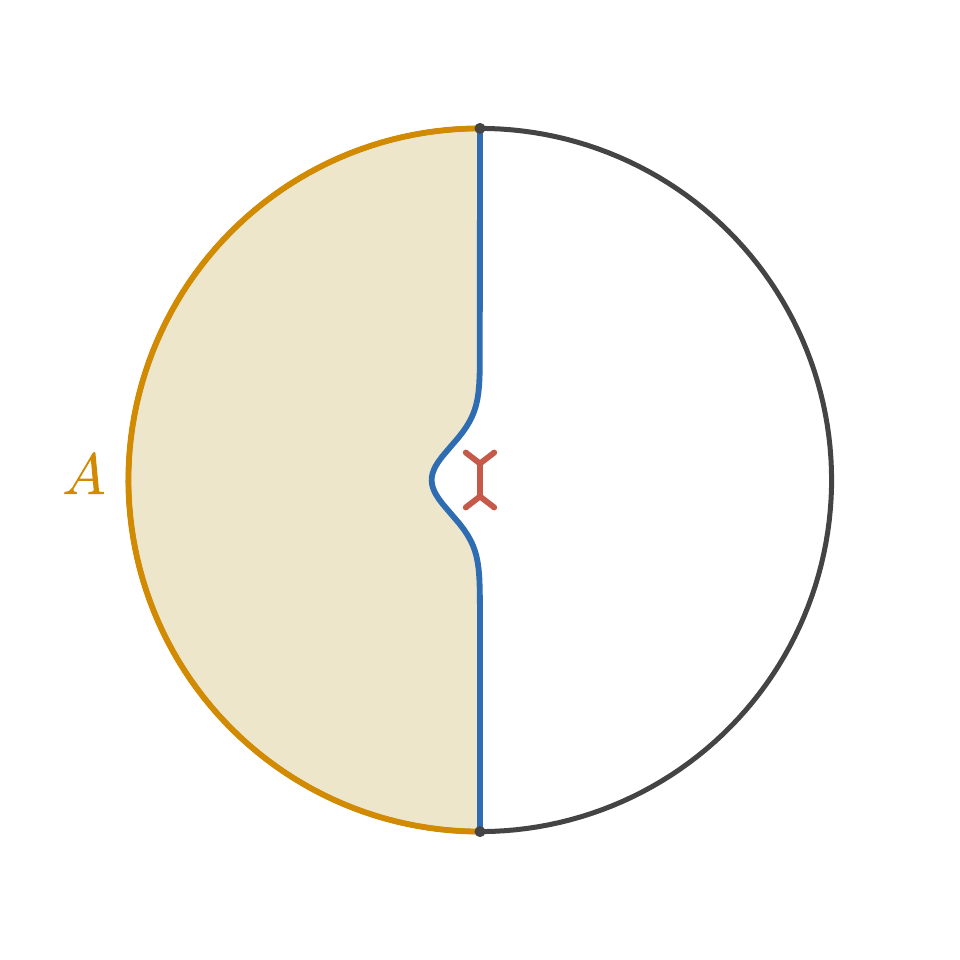}{\linewidth}{(a)}
    \end{minipage}
    \begin{minipage}[b]{0.48\linewidth}
        \panel{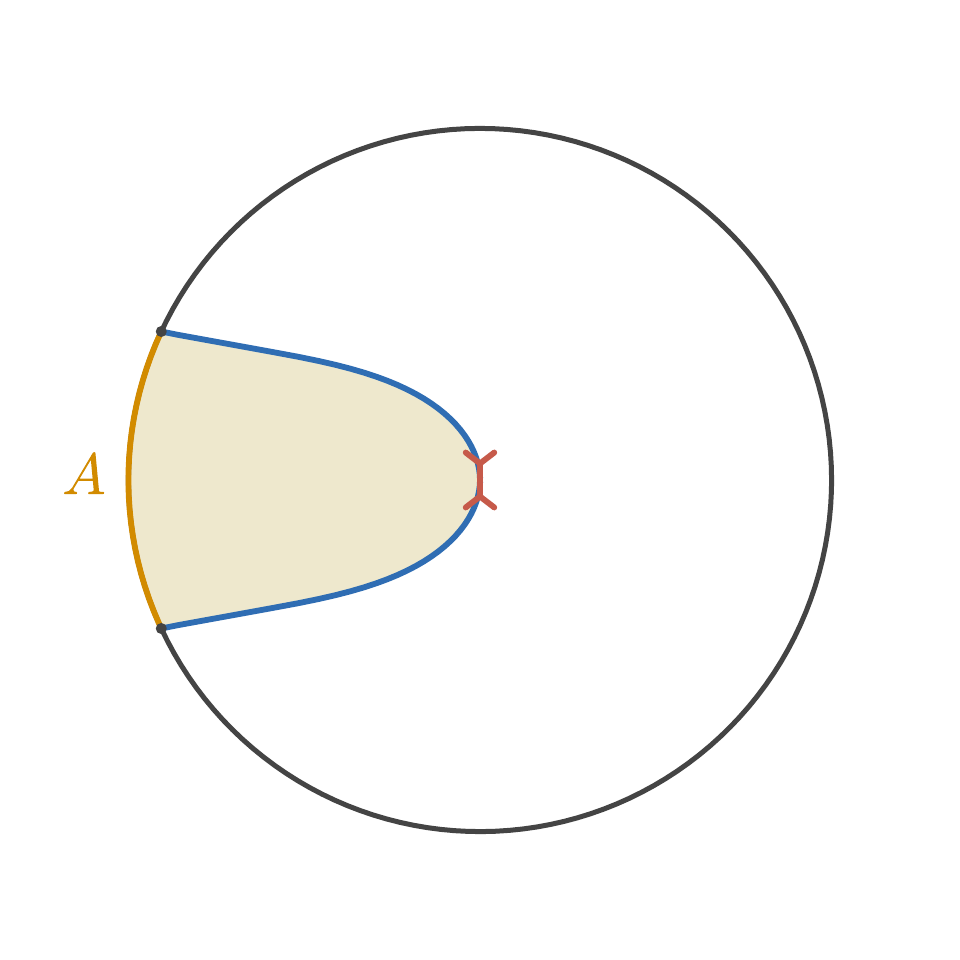}{\linewidth}{(b)}
    \end{minipage}
	\caption{Extremal surfaces in the emergent geometries of finite-depth circuit. (a) In AdS, the boundary-to-boundary RT geodesic turns around before reaching the IR state. (b) In dS spacetime, the timelike extremal geodesics extend to the IR endpoint, suggesting that the entanglement retained in the remaining IR state may affect the entanglement entropy.}
    \label{fig:disk}
    \end{figure}

The contrast between AdS and dS extremal curves provides a geometric picture for sensitivity to infrared information. In AdS, the boundary-to-boundary RT geodesic turns around before reaching the IR endpoint, as shown in Fig.~\ref{fig:disk}(a). The entanglement entropy thus need not probe the IR degrees of freedom associated with scales below the finite-size scale, consistent with the insensitivity of Hermitian critical systems to a gap $\Delta<1/L$. In dS, by contrast, the timelike extremal curves extend to the IR endpoint for every boundary interval, as shown in Fig.~\ref{fig:disk}(b). This distinction suggests why the entanglement entropy can remain sensitive to the degrees of freedom retained at the end of the RG flow, even for gaps below the finite-size scale. We now examine the entropy of the remaining two-site state directly.

\emph{Residual entropy in the IR state.}---We first estimate the IR contribution from the cMERA phase rotation left unperformed at finite depth, and then compute the IR-state entropy exactly.  The geometric length associated with the remaining
rotation is identified with the residual change of the band phase,
\begin{equation}
2\int ds
=
2\int\sqrt{g_{uu}}\,du
=
\int du\,\partial_u\phi_k(u)
=
\Delta\phi_{\rm res}.
\label{eq:geo_phase_id}
\end{equation}
On a ring, the circuit disentangles modes only down to \(k_{\min}=2\pi/L\), whereas the fully disentangled reference requires continuing to \(k=0\).  
Since \(\phi(k)\sim i\log k\) near the EP, with the \(k=0\) singularity cut off by \(\Delta\), the remaining rotation is
\begin{equation}
\Delta\phi_{\rm res}
=
\phi_{k=0}-\phi_{k=2\pi/L}
=
i\log\!\left(\frac{\Delta L}{2\pi}\right).
\label{eq:residual_phase_scaling}
\end{equation}
The remaining phase rotation therefore exhibits the same logarithmic dependence on \(\Delta L\) as the residual entropy.

We now calculate the entropy of the two-site IR state directly from its correlation matrix.
This two-site IR endpoint is also consistent with the periodic dS/MERA tensor network of Ref.~\cite{chou2026emergentsitterspacenonunitary}.
In momentum space, it is represented by the two modes \(k=0\) and \(k=\pi\).
The finite-depth circuit has already implemented the rotation from \(k=\pi\) to \(k=2\pi/L\), leaving
\begin{equation}
\phi'(\pi)=\phi(\pi),
\qquad
\phi'(0)
=
\phi(0)-\big[\phi(2\pi/L)-\phi(\pi)\big].
\label{eq:phi_prime_IR}
\end{equation}
Thus, the phase difference stored in the IR state is precisely the missing \(k=0\) to \(k=2\pi/L\) rotation in Eq.~\eqref{eq:residual_phase_scaling}.

\begin{figure}
    \begin{minipage}[b]{0.97\linewidth}
        \panel{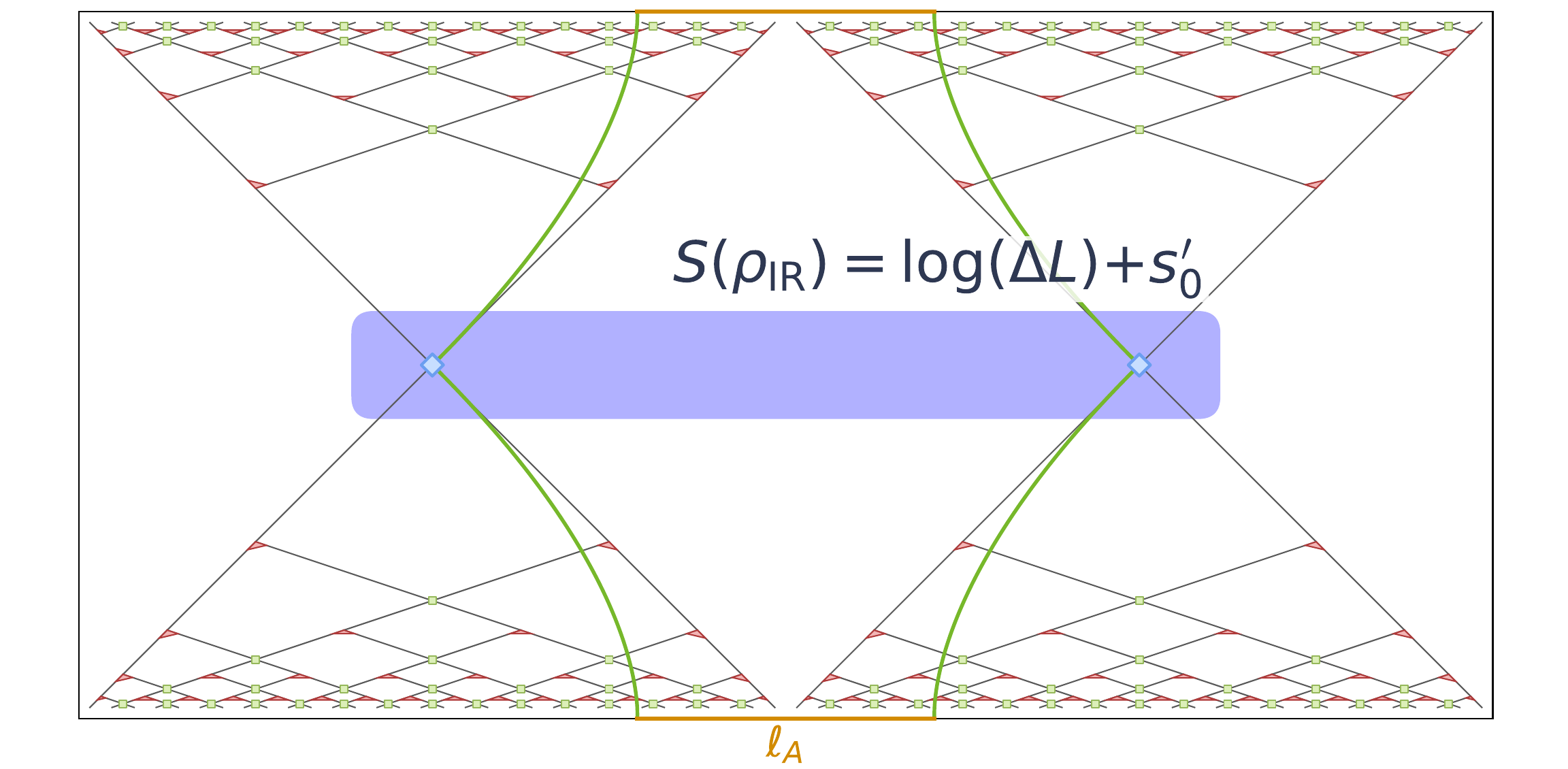}{\linewidth}{(a)}
    \end{minipage}
    \begin{minipage}[b]{0.48\linewidth}
        \panel{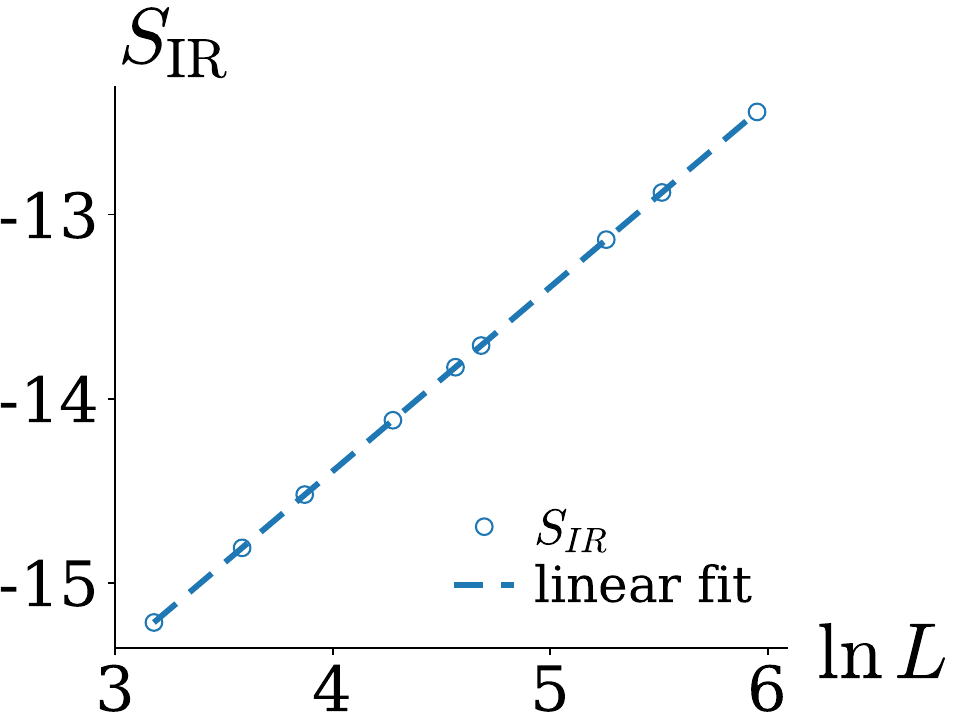}{\linewidth}{(b)}
    \end{minipage}
    \begin{minipage}[b]{0.48\linewidth}
        \panel{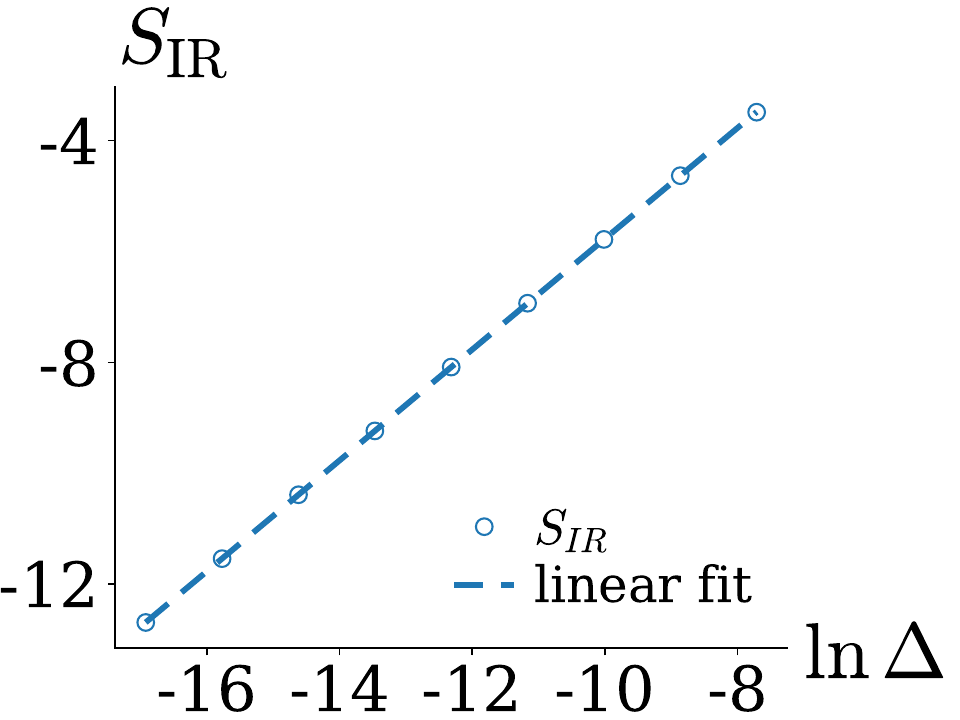}{\linewidth}{(c)}
    \end{minipage}
	\caption{Residual entropy of the two-site IR state.
    (a) In the periodic dS/MERA network, extremal curves associated with different boundary intervals follow different bulk paths but terminate at the same pair of IR endpoints, providing a geometric picture for the interval-independent contribution.
    (b) \(S_{\rm IR}\) versus \(\log L\) at fixed
    \(\Delta=10^{-8}\).
    (c) \(S_{\rm IR}\) versus \(\log\Delta\) at fixed \(L=100\).
    For (b) and (c), we set \((v,w)=(2,1)\); both linear fits give a
    slope \(0.999\), confirming
    \(S_{\rm IR}=\log(\Delta L)+s_0^\prime\).}
    \label{fig:IR_entropy}
\end{figure}

Evaluating the two-site correlation matrix gives
\begin{equation}
S_{\rm IR}
=
\log(\Delta L)
+
s_0^\prime.
\label{eq:IR_entropy_result}
\end{equation}
An asymptotic analytic derivation of Eq.~\eqref{eq:IR_entropy_result} is given in SM~\cite{SM}.
The unit-slope dependences on \(\log\Delta\) at fixed \(L\) and on \(\log L\) at fixed \(\Delta\) are shown in Fig.~\ref{fig:IR_entropy}(b) and (c).  
The IR-state entropy therefore reproduces the additional term in Eq.~\eqref{eq:intro_decomp}, identifying the residual entropy as the entropy carried by the entanglement remaining after finite-depth disentangling.

Finally, the periodic dS/MERA structure provides a natural geometric picture for the interval independence of the residual entropy. As shown in Fig.~\ref{fig:IR_entropy}(a), extremal curves associated with different boundary intervals follow different bulk paths but terminate at the same IR endpoints. This common endpoint suggests how the remaining IR state can contribute equally to the entanglement entropy of intervals of different sizes.

\emph{Conclusion.}---In this work, we have shown that non-Hermitian critical free fermions near exceptional points remain sensitive to a small energy gap \(\Delta\), even when \(\Delta<1/L\). For a subsystem of length \(\ell_A\) on a ring of length \(L\), the usual Calabrese–Cardy dependence is supplemented by an additional \(\ell_A\)-independent \(\log(\Delta L)\) term when \(\Delta L\lesssim 1\). Such sensitivity to gaps below the finite-size scale is absent in the corresponding unitary systems.

To investigate the infrared origin of this contribution, we considered a finite-depth non-unitary RG circuit. Completing the flow to a fully disentangled one-site state produces a singular final layer, while terminating the circuit one layer earlier leaves a two-site IR state. Its entanglement entropy reproduces the same \(\log(\Delta L)\) dependence found in the full lattice calculation, supporting the interpretation of the residual entropy as entanglement retained after finite-depth disentangling.

The emergent dS geometry provides a complementary interpretation. Unlike the AdS RT geodesics, the dS extremal curves extend to the IR endpoint for every boundary interval, suggesting why the entanglement entropy can remain sensitive to infrared information. Moreover, curves associated with different intervals terminate at the same IR endpoints, providing a natural geometric picture for the interval independence of the additional contribution.

Our results show that logarithmic non-Hermitian critical free fermions can exhibit an additional entanglement contribution beyond the standard Calabrese–Cardy scaling. The two-site IR-state calculation supports its infrared origin, while the emergent de Sitter geometry provides a natural picture for its sensitivity to infrared information and independence of subsystem size.
Looking forward, it would be important to understand whether this residual entropy is a general feature of logarithmic non-unitary criticality, or a peculiarity of the free-fermion models studied here.  
One natural direction is to derive this additional term directly from logarithmic CFT, rather than only from a lattice calculation.   
Another important question is whether the residual entropy survives interactions.
From the holographic side, a more systematic dS/LCFT correspondence may clarify its bulk interpretation and physical meaning.

\emph{Acknowledgement:} The author is grateful to Po-Yao Chang for many stimulating discussions and valuable suggestions.
This work was supported by the National Science and Technology Council of Taiwan under Grant No.~NSTC 114-2112-M-007-015.

\bibliographystyle{apsrev4-2}
\bibliography{reference}

\onecolumngrid

\clearpage
\newpage

\widetext

\begin{center}
\textbf{\large Supplemental Material for ``Beyond Calabrese–Cardy Scaling: Exceptional-Point Sensitivity and the de Sitter RT Surface''}
\end{center}


\renewcommand{\thefigure}{S\arabic{figure}}
\setcounter{figure}{0}
\renewcommand{\theequation}{S\arabic{equation}}
\setcounter{equation}{0}
\renewcommand{\thesection}{\Roman{section}}
\setcounter{section}{0}
\setcounter{secnumdepth}{4}
\counterwithout*{equation}{section}

\section{Residual entropy in additional free-fermion models}

In the main text, we demonstrated the residual entropy using the two-legged
non-Hermitian SSH model. To examine whether the same behavior depends on the
particular lattice realization, we first consider the non-Hermitian SSH model
\begin{equation}
H_{\rm SSH}(k)=
\begin{pmatrix}
iu & v_k\\
v_k^* & -iu
\end{pmatrix},
\qquad
v_k=v-we^{-ik},
\label{eq:sm_nhssh}
\end{equation}
where $u,v,w\in\mathbb{R}$. Its spectrum is
\begin{equation}
E_\pm(k)=\pm\sqrt{|v_k|^2-u^2}.
\label{eq:sm_nhssh_spectrum}
\end{equation}
The gap closes at $k=0$ when $|v-w|=u$, where the Bloch Hamiltonian becomes
defective. The cases $v>w$ and $w>v$ correspond to the trivial and topological
critical points, respectively. The two-legged model considered in the main text
instead has $v_k=v-w\cos k$. In all cases, we approach the exceptional point
from the gapped regime and denote the energy gap by $\Delta$. 
For the SSH numerical calculations, we fix $(v,w)$ and set \(u=|v-w|-\delta\), with \(\delta>0\), giving the physical gap \(\Delta=\sqrt{2|v-w|\delta-\delta^2}\).

We also consider the symplectic-fermion model~\cite{GAINUTDINOV2013245,KAUSCH2000513},
\begin{equation}
H_{\rm SF}(k)=
\begin{pmatrix}
m & q(k)\\
-q^*(k) & -m
\end{pmatrix},
\qquad
q(k)=i(1+e^{ik}),
\label{eq:sm_sf}
\end{equation}
with spectrum
\begin{equation}
E_\pm(k)=\pm\sqrt{m^2-|q(k)|^2}.
\label{eq:sm_sf_spectrum}
\end{equation}
At $m=2$, the gap closes at the exceptional point $k=0$. In the numerical
calculation, we take $m=2+\delta$ with $\delta>0$ to open a small energy gap.
We further consider the two-legged version, obtained by replacing
$q(k)$ with $i(1+\cos k)$.

For each model, the entanglement entropy is computed from the biorthogonal
correlation matrix.  The right and left single-particle eigenstates satisfy
\begin{equation}
H|R_\alpha\rangle=E_\alpha|R_\alpha\rangle,
\qquad
H^\dagger|L_\alpha\rangle=E_\alpha^*|L_\alpha\rangle,
\qquad
\langle L_\alpha|R_\beta\rangle=\delta_{\alpha\beta}.
\end{equation}
At half filling, the corresponding many-body ground states are denoted by
$|G_R\rangle$ and $\langle G_L|$, with
$\langle G_L|G_R\rangle=1$, and the biorthogonal density matrix is
\begin{equation}
\rho=|G_R\rangle\langle G_L|.
\end{equation}
For free fermions, the entanglement entropy can be obtained directly from the
restricted correlation matrix~\cite{IngoPeschel_2003,PhysRevResearch.2.033069}. For a subsystem
$A$, the restricted correlation matrix is
\begin{equation}
(C_A)_{ij}
=
\sum_{\alpha\in{\rm occ}}
\langle L_\alpha|i\rangle
\langle j|R_\alpha\rangle,
\qquad i,j\in A,
\label{sm_correlation}
\end{equation}
and its eigenvalues $\{\nu_n\}$ determine the von Neumann entropy,
\begin{equation}
S_A=
-\sum_n
\left[
\nu_n\log\nu_n+(1-\nu_n)\log(1-\nu_n)
\right].
\label{eq:sm_entropy}
\end{equation}
Unlike a Hermitian correlation matrix, $C_A$ is generally non-Hermitian.
Its eigenvalues are therefore not restricted to the interval $[0,1]$ and can
take negative values or become complex, allowing the biorthogonal entanglement
entropy to be negative.

\begin{figure}
    \begin{minipage}[b]{0.32\linewidth}
        \panel{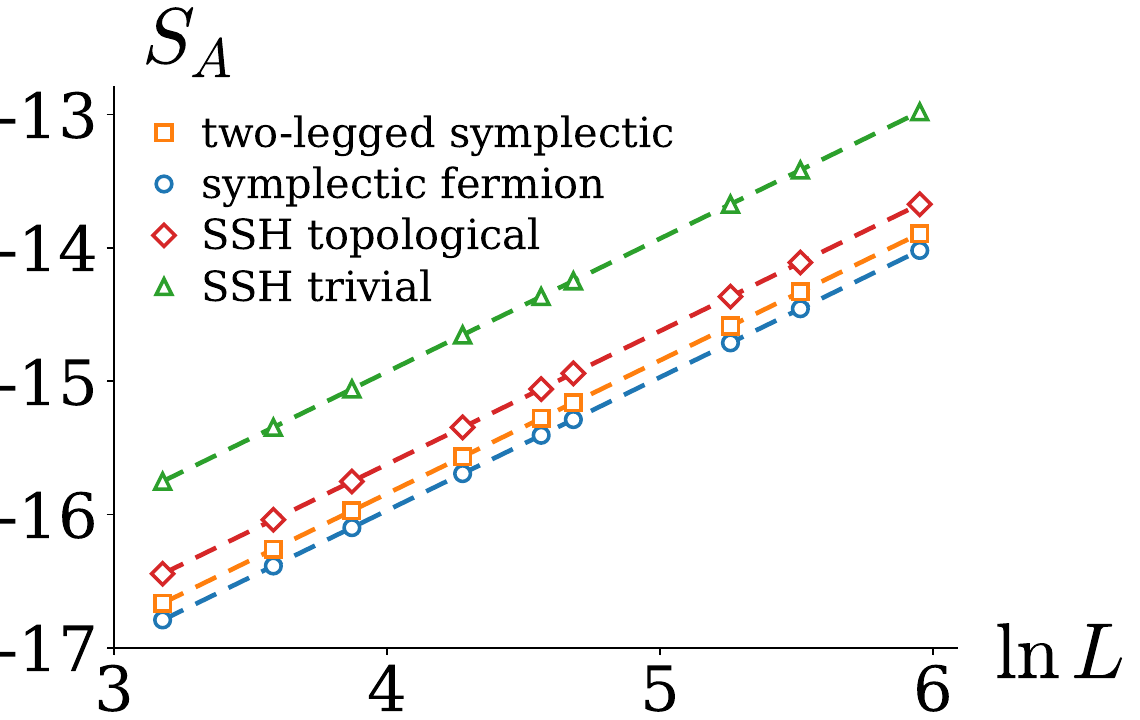}{\linewidth}{(a)}
    \end{minipage}
    \begin{minipage}[b]{0.32\linewidth}
        \panel{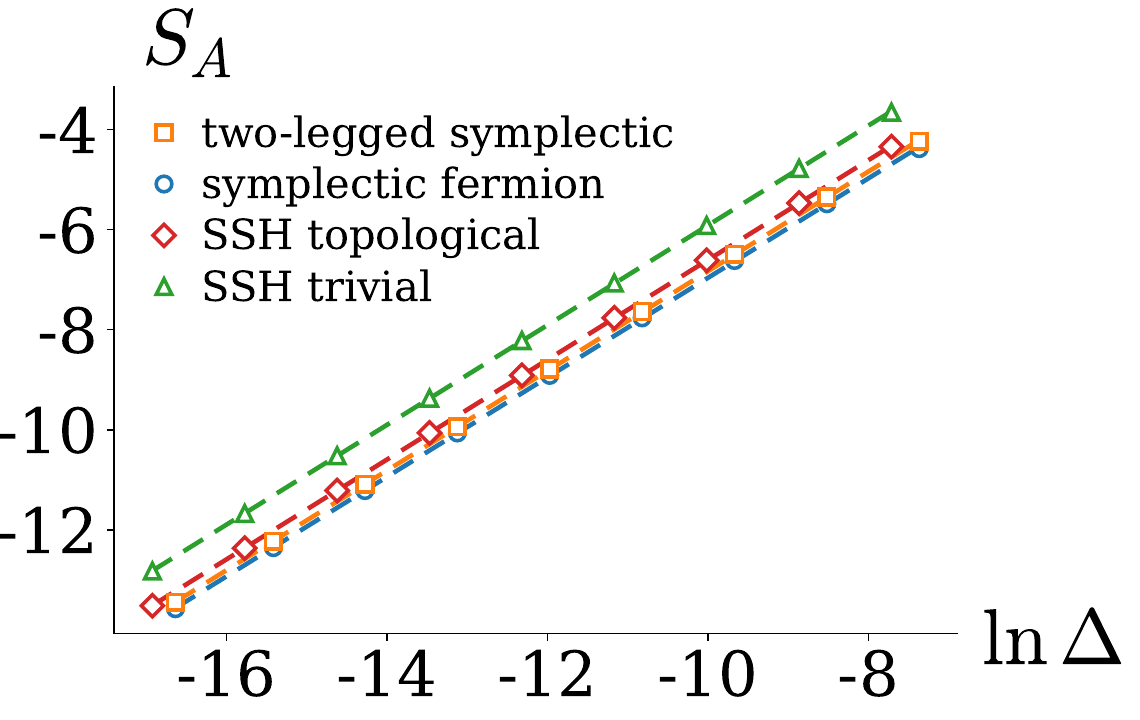}{\linewidth}{(b)}
    \end{minipage}
    \begin{minipage}[b]{0.32\linewidth}
        \panel{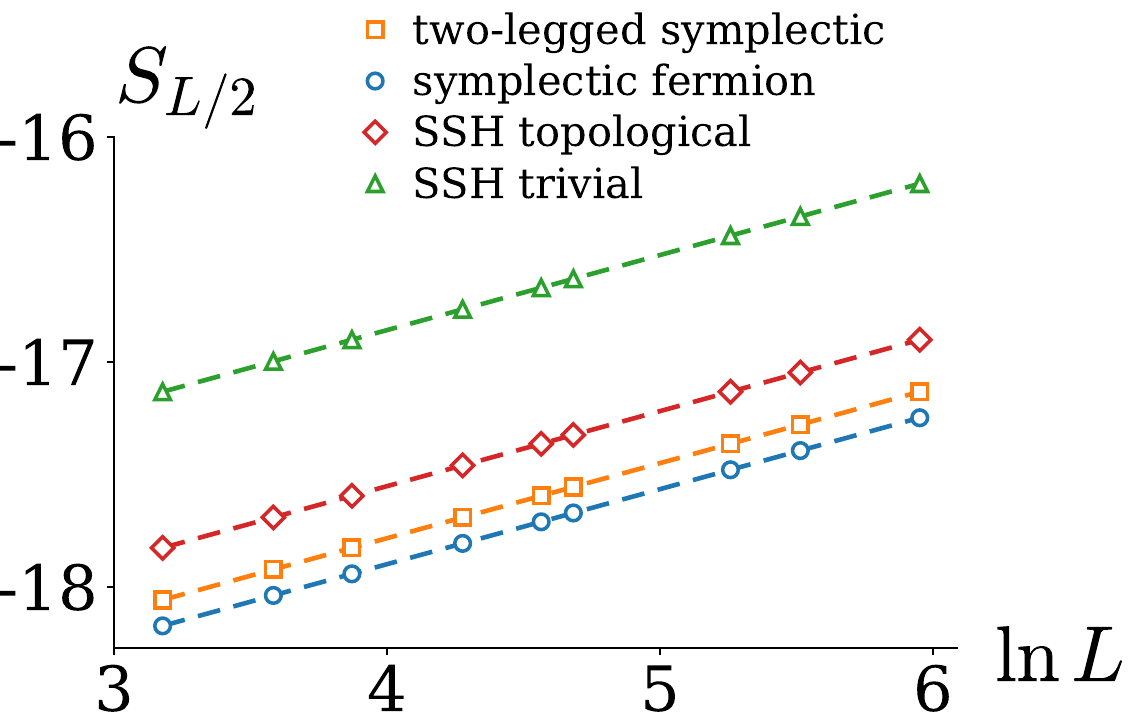}{\linewidth}{(c)}
    \end{minipage}
	\caption{
Residual-entropy scaling in four one-dimensional non-Hermitian
free-fermion models: the symplectic-fermion model, its two-legged
realization, and the non-Hermitian SSH model at the trivial and
topological critical points.
For the symplectic-fermion models, we use the parameters defined in
the text with unit hopping, while for the trivial and topological SSH
critical points we take $(v,w)=(2,1)$ and $(1,2)$, respectively.
(a) Entanglement entropy of a fixed one-site subsystem as a function
of $\log L$ at fixed physical gap $\Delta=10^{-8}$.
The fitted slopes are $1.000$, $1.000$, $1.000$, and $1.000$ for the
symplectic-fermion, two-legged symplectic-fermion, trivial SSH, and
topological SSH models, respectively, consistent with
$S_A=\log L+\mathrm{const.}$
(b) One-site entanglement entropy as a function of $\log\Delta$ at
fixed system size $L=100$.
The corresponding fitted slopes are $0.994$, $0.996$, $0.997$, and
$0.997$, confirming $S_A=\log\Delta+\mathrm{const.}$
(c) Half-chain entropy as a function of $\log L$ at fixed physical
gap $\Delta=10^{-8}$.
The fitted slopes are $0.333$, $0.333$, $0.333$, and $0.333$, in
agreement with
$S_{L/2}=\frac{1}{3}\log L+\mathrm{const.}$
for $c=-2$ after including the residual contribution.
Dashed lines denote linear fits to the numerical data.
}
    \label{fig:residual_other_models}
\end{figure}

The results are summarized in Fig.~\ref{fig:residual_other_models}. For a fixed
one-site subsystem, $\ell_A=1$, the Calabrese--Cardy contribution is independent
of $L$, and the entropy scales as
\begin{equation}
S_A(\ell_A=1)=\log L+\mathrm{const.}
\label{eq:sm_logL}
\end{equation}
at fixed $\Delta$, as shown in Fig.~\ref{fig:residual_other_models}(a).

At fixed $L$ and subsystem size, the entropy instead scales as
\begin{equation}
S_A=\log\Delta+\mathrm{const.},
\label{eq:sm_logDelta}
\end{equation}
as shown in Fig.~\ref{fig:residual_other_models}(b). Together,
Eqs.~\eqref{eq:sm_logL} and~\eqref{eq:sm_logDelta} give
\begin{equation}
S_{\rm res}=\log(\Delta L)+\mathrm{const.}
\label{eq:sm_residual}
\end{equation}
for all the models considered here.

Finally, for a half-chain subsystem, $\ell_A=L/2$, combining
the residual contribution in Eq.~\eqref{eq:sm_residual} with the $c=-2$
Calabrese--Cardy term gives
\begin{equation}
S_A\left(\frac{L}{2}\right)
=
\frac{1}{3}\log L+\mathrm{const.}
\label{eq:sm_halfchain}
\end{equation}
at fixed $\Delta$, as shown in Fig.~\ref{fig:residual_other_models}(c).

These results show that the coefficient of the residual logarithm is unchanged
between the non-Hermitian SSH and symplectic-fermion models, and between the
trivial and topological SSH critical points.

\section{Asymptotic derivation of the two-site IR entropy}
\label{app:IR_asymptotic}

We derive the asymptotic two-site result \(S_{\rm IR}=\log(\Delta L)+\mathrm{const.}\) stated in the main text.  After the finite-depth flow, the remaining IR sector contains only the \(k=0,\pi\) modes.  Writing their single-particle projectors as
\begin{equation}
P_k=\ket{R_k}\bra{L_k},
\qquad
\ket{R_k}
=
\begin{pmatrix}
-\sin(\phi_k/2)\\
\cos(\phi_k/2)
\end{pmatrix},
\qquad
\bra{L_k}
=
\begin{pmatrix}
-\sin(\phi_k/2)&\cos(\phi_k/2)
\end{pmatrix}.
\end{equation}
Using Eq.~\eqref{sm_correlation}, the correlation matrix for either site of the
two-site system is
\begin{equation}
C_A=\frac{1}{2}\left(P_0+P_\pi\right).
\label{eq:IR_CA}
\end{equation}
Diagonalizing Eq.~\eqref{eq:IR_CA} gives
\begin{equation}
\nu_\pm
=
\frac12
\pm
\frac12
\cos\!\left(\frac{\phi_0'-\phi_\pi'}{2}\right).
\label{eq:IR_nu_exact}
\end{equation}
The finite-depth circuit leaves the \(k=0\) to \(k=2\pi/L\) rotation unperformed, so in the regime \(\Delta L\ll1\),
\begin{equation}
\phi_0'-\phi_\pi'
\simeq
i\log\!\left(\frac{\Delta L}{2\pi}\right).
\label{eq:IR_phase_asymptotic}
\end{equation}
Substituting Eq.~\eqref{eq:IR_phase_asymptotic} into Eq.~\eqref{eq:IR_nu_exact} gives
\begin{equation}
\nu_\pm
=
\frac12
\pm
\frac12
\cosh\!\left[
\frac12\log\!\left(\frac{\Delta L}{2\pi}\right)
\right].
\label{eq:IR_nu_asymptotic}
\end{equation}
Substituting Eq.~\eqref{eq:IR_nu_asymptotic} into the entropy formula,
Eq.~\eqref{eq:sm_entropy}, and expanding for $\Delta L\ll1$ gives
\begin{equation}
S_{\rm IR}
=
\log(\Delta L)+s_0'+O(\Delta L),
\label{eq:IR_entropy_scaling}
\end{equation}
where \(s_0'\) is an \(O(1)\) constant.

The two-site IR state therefore reproduces the unit-coefficient \(\log(\Delta L)\) residual contribution found in the full lattice entropy.

\end{document}